# Resonant Raman of OH/OD vibrations and photoluminescence studies in LiTaO$_3$ thin film

S. Satapathy[a], B. N. Raja Sekhar[b], V. G. Sathe[c], Shailendra Kumar[a]*

and P. K. Gupta[a]

[a]Laser Materials Development & Devices Division, Raja Ramanna Centre for Advanced Technology, Indore 452013, India

[b]Spectroscopy Division, Bhabha Atomic Research Centre, Mumbai 40085, India

[c]UGC-DAE Consortium for Scientific Research, Indore 452017, India

E-mail address: Shailendra Kumar  -  shail@cat.ernet.in

Phone : +91-731-2488377

FAX: +91-731-2488300

**Abstract**

Resonant Raman spectra (RRS) of O-H and O-D vibration and libration modes, their combinations and higher harmonics have been observed in $LiTaO_3$ polycrystalline thin films. RRS peaks are superimposed on photoluminescence (PL) spectrum. Monochromatic light from a xenon lamp is used as excitation source. PL spectrum shows two broad peaks, first near the band gap in UV (4.4-4.8eV) and another in the sub band gap region (< 4.0 eV). Band gap PL along with RRS peaks are reported for the first time. Photoluminescence excitation spectrum (PLE) shows a peak at 4.8 eV. Peak positions and full width at half maximum (FWHM) of RRS peaks depend upon the excitation energy. Dispersions of the fundamental and the third harmonic of the stretching mode of O-H with excitation energy are about 800 $cm^{-1}$/eV and 2000 $cm^{-1}$/eV respectively. This dispersion is much higher than reported in any other material.

1. Introduction

Lithium tantalate (LiTaO$_3$) is one of the widely used optoelectronic material because of its ferroelectric, piezoelectric and pyroelectric properties. It is also an attractive material for integrated–optic applications due to its nonlinear optical properties, large electroptic and piezoelectric coefficients [1] and its superior resistance to laser induced optical damage [2]. Now days LiTaO$_3$ (LT) thin film has got importance because of its potential uses in waveguide structures. By combining ferro-electric thin films of LT with integrated semiconductor chips, a new unique family of devices can be fabricated. Pyroelectric detectors using LT thin film, deposited by sol-gel method, have been demonstrated [3].

Presence of O-H had been reported in as grown LT and LiNbO$_3$ (LN) crystals, observed by IR and Raman spectroscopies by many groups [4-6]. Stretching mode (SM) and libration mode (LM) of O-H / O-D have been reported in protron / deutron exchanged LT crystals using FTIR spectroscopy and magnetic resonance spectroscopy [4-6]. Presence of O-D is much less than O-H in as grown LT and LN crystals and it is difficult to detect it using FTIR. Observation of O-H and O-D vibration modes by resonance Raman spectroscopy (RRS) along with photoluminescence (PL) in the U-V range has not been reported earlier. In RRS, incident or outgoing energy is in resonant with allowed electronic transitions. Theoretical models of RRS had been published in early seventies [7, 8]. RRS had been used extensively in semiconductors, carbon compounds and nanotubes [9-17]. In the present work, RRS vibration and libration modes of O-H and O-D in as grown LT thin films along with PL spectra in UV and visible region are reported at room temperature. Further, it has been observed that the RRS peak positions depend upon the excitation energy. The observed dispersion with excitation energy for fundamental stretching mode of O-H is about 800 cm$^{-1}$/eV and for its third harmonic it is about 2000cm$^{-1}$/eV, which are much larger than the dispersion reported in any other material.

**2. Experiment**

LT thin films were deposited on Pt-Si [Pt(111)/SiO$_2$/TiO$_2$/Si(100)] substrate by spin coating of sol gel and details of the film preparation and structural characterizations are given in an earlier paper [18]. Thin films are characterized using X-ray diffraction (XRD) and atomic force microscopy (AFM). Raman and photoluminescence spectra were recorded on polycrystalline LT thin films at room temperature to find out the possible Raman modes and understand the nature of the film. The Raman measurements on LT films were carried out using 488 nm excitation source using LABRAM-HR spectrometer equipped with a Peltier cooled charge coupled device (CCD) detector. The typical spectral resolution was ~1cm$^{-1}$. PL studies were carried out using unpolarized excitation source as the incident ray. Emitted rays were recorded using back reflection geometry and spectra were recorded with incremental step of 0.1 nm. The photoluminescence and resonance Raman spectral data were obtained using a spectrofluorometer (FluoroMax –3, Jobin Vyon). The background excitation source used for this purpose was a 150W xenon, continuous output, ozone-free lamp. A calibrated photodiode and a R928P photomultiplier are used as excitation and emission detectors.

## 3. Results and Discussion

Polycrystalline thin films have been deposited without any preferential orientation have been confirmed by grazing angle X-ray diffraction spectra. The root mean square roughness of the thin film is 3.9 nm (over an area of 3μm by 3μm) for film of thickness~400nm. The thickness of thin films is around 400nm measured using filmatrics instrument in the reflection mode.

**Photoluminescence Studies**

Photoluminescence excitation (PLE) spectra as a function of incident wavelengths of thin film were recorded by keeping the detector at 280nm. Incident wavelengths were varied from 225nm to 275nm. PLE spectrum consists of sharp peak at 259nm (4.8eV) and a broad peak due to band gap (fig.1). The peak at 259nm was also observed, when detector was kept at 263nm. Inbar and Cohen calculated the electronic structure of LT [19]. The top of the valence band is between Γ and Z, and the bottom of conduction band is at Γ point. The top of the valence band is almost flat between the Γ point and the mid point of Γ and Z line [19]. The lowest energy indirect transition between the bottom of the conduction band and the top of the valence band is 4.4eV and the next higher allowed transition between energy levels at Γ point is around 4.8eV, according to the reported energy-momentum diagram. In the present report the sharp peak observed at 259 nm in PLE spectra is due to the allowed transition at 4.8eV. This observed value matches with calculated E-k diagram [19].

The PL spectra at room temperature of LT thin film for two excitation wavelengths 235nm and 240nm are shown in fig. 2(a). The incident light was unpolarized. Two broad emission bands were detected. The broad PL emission peak between 255nm and 285nm corresponds to band gap of LT thin film. The broad PL band in the range 300nm-445nm, (sub band gap of LT ) centred around 370 nm is observed due to defect centres connected with a great number of oxygen vacancies and matches with published PL spectra [20, 21]. The band gap PL in the UV range has not been reported earlier either in thin films or single crystals of LT. If LT is an indirect band gap material, then question arises that what is the source of the band gap PL ? The structure of the LT thin film is strained near the interface with Pt/Si substrate [18]. The band gap of the strained LT thin film may be direct and hence, the band gap PL in the UV range is observed in the present case. Further, the variation in the strain in the structure will give broad band gap and hence the broad PL peak in the range (4.27eV to 4.96eV) 255nm to 285nm. As reported, by Kim et.al., the band gap of LT crystals depend upon the Li concentration and the absorption edge varies from 272nm to 282nm when $Li_2O$ mole percentage changes from 50% to 45% [22]. The variation of the Li concentration is not expected in a LT thin film and it may vary from one deposition to another. Finally, strain seems to be the main cause of the band gap PL from LT thin films. Fig. 2(b) shows PL spectra plotted with X-axis as shift of energy with respect to excitation energy. Small peaks are superimposed on PL spectra are discussed in the next section.

**Raman and Resonance Raman**

The ferroelectric phase of LT belongs to R3C space group and two molecules per unit cell. Out of total 30 degrees of freedom 4 $A_1$ (z) and 18E [9E(x) + 9E(y)] modes are Raman and infrared active [23-26]. The fundamental phonon mode positions in Raman spectrum reported in the literature are 74, 140, 206, 251, 316, 383, 462, 596 and 662$cm^{-1}$

for E (TO); 80, 163, 248, 278, 318, 452, 474, 648 and 870 cm$^{-1}$ for E (LO); 201, 253, 356 and 600 cm$^{-1}$ for A$_1$ (TO) and 245, 347, 401 and 864 cm$^{-1}$ for A$_1$ (LO) [23-26].

Fig. 3 shows normal Raman spectra of LT thin films, excited with 488nm light. The spectrum shows many sharp and broad lines at various phonon modes and oblique phonon modes. The major peaks appear at 143, 207, 250, 310, 353, 374, 463, 593, 657 and 858cm$^{-1}$. Broad peaks are due to superposition of more than one phonon modes. All major peaks have been assigned to different fundamental modes as shown in fig. 3 and these match with the reported phonon modes. The spectrum shows peaks for both A$_1$ and E modes simultaneously due to polycrystalinity of thin film. The minor peaks, other than major peaks, may be attributed to the strain at substrate and thin film interface. Further, lattice disorder due to oxygen vacancies will lower the symmetry of the material and contribute to minor peaks.

Small peaks superimposed on PL spectra are indicated by dashed lines in fig. 2(b). Peak positions for 235nm excitation are shifted to higher energy with respect to 240nm excitation. Fig. 4 shows the PL spectra in the band gap region measured by varying the incident angle between 42$^0$ and 53$^0$ for the excitation wavelength 240nm. Observed superimposed peak positions for 240nm excitation in fig. 2(b) and fig.4 are listed in Table 1 with possible modes assigned to these peaks. The incident angle was varied to maximize the intensity of PL and resonance Raman (RR) peaks, as most of the time the front surface of the sample and the holder are not exactly parallel to each other. As shown in figure 4, the intensity of some modes is angle dependent. Experiment was repeated with polarized incident light, but the intensity of PL and RR peaks was too low to do meaningful measurement. For a meaningful answer for this observation, further studies are planned using intense polarised excitation source such as synchrotron light source.

Grone and Kaphan had observed vibration and libration modes of proton and deutron exchanged $LiTaO_3$ crystals using FTIR spectra. Their results are also listed in Table1. Vibrational modes related with O-H in $LiTaO_3$ crystals had been reported by many groups [4, 27]. Vibrational modes related with O-D in $LiTaO_3$ crystals (without deuteron exchange) had not been reported. In the present observation, the FWHM of O-H and O-D vibrational modes in RRS spectra (fig 4) are larger by a factor of 10 in comparison to reported FTIR spectra. RRS peaks observed for 240nm excitation are discussed below. Stretching vibrations of O-H and O-D are observed at 3637 $cm^{-1}$ and 2535 $cm^{-1}$ respectively. O-D stretching mode was observed in two samples out of five samples studied. O-H stretching mode and higher harmonics were observed in all five samples. As, about 0.015% $D_2O$ is present in water there is probability that we could see the presence of O-D modes in few samples at resonance condition. Further, Raman spectrum was also measured in range 2000-6000 $cm^{-1}$, excited with 488nm laser light, and RR peaks related with O-H and O-D stretching modes were not detectable. In fig 4, RR Peak at 2764 $cm^{-1}$ is due to stretching mode – libration mode of O-H and peak at 4558 $cm^{-1}$ is due to stretching mode + libration mode of O-H. Peak at 5150 $cm^{-1}$ is 2$^{nd}$ harmonic of stretching mode of O-D. Peak near 5975 $cm^{-1}$ with FWHM of about 350 $cm^{-1}$ is superimposition of two peaks. First is due to combination of stretching modes of O-H and O-D and another due to combination of 2$^{nd}$ harmonic of stretching mode of O-D with libration mode of O-D. Peak at 6350 $cm^{-1}$ is due to difference of 2$^{nd}$ harmonic of stretching mode of O-H and libration mode of O-H. Peak at 6630 $cm^{-1}$ is due to 2$^{nd}$ harmonic of combination of stretching with libration modes of O-D. Peak at 11040 cm$^{-1}$ is due to 3$^{rd}$ harmonic of stretching mode of O-H. Peak at 13100 cm$^{-1}$ is due to combination of 3$^{rd}$ harmonic of stretching mode of O-H with fundamental stretching mode of O-D. Peak at 15600 cm$^{-1}$ is due to combination of 3$^{rd}$ harmonic of stretching mode of O-H and 2$^{nd}$ harmonic of stretching mode of O-D. Main observations are that measurable peaks due to 2$^{nd}$ harmonic of O-H and 3$^{rd}$ harmonic of O-D have not been observed.

RRS peak positions depend upon the excitation energy. Fig 5(a) shows RRS peak positions of the fundamental stretching mode of O-H (for constant incident photon flux) for different excitation wavelengths. Excitation wavelengths are shown near respective peaks. Fig 5 (b) shows variation of the RRS peak intensity (for constant incident photon flux) as a function of incident energy. In fig 5(b), there is a broad maximum in the range 4.6-4.8 eV, a minimum near 4.96 eV and peak intensity increases for E > 5 eV. Further, energies of minima in fig 5(b) and fig 1 are same. In fig1, the intensity of PL reduces for E > 5eV, due to increase of surface recombination of excess carriers. The increase of RR peak intensity for E > 5eV, in fig 5(b), is interesting and needs further studies. Dispersions of the fundamental stretching modes of O-H and its third harmonic with excitation energy are shown in fig 5(c) by curves 1 and 2 respectively. The slope of the dispersion curve-1 in fig.5(c) is 812 cm$^{-1}$/eV and the slope for the curve-2 in fig.5(c) is 2210 cm$^{-1}$/eV. These dispersions with excitation energy are much larger than reported in any other material. Multi wavelengths Raman studies of the D band in amorphous carbon had shown dispersion with excitation wavelengths and it is about 50cm$^{-1}$/eV [16,17]. For excitation wavelength 250nm, the RRS peak positions for fundamental stretching mode of O-H ( 3450cm$^{-1}$) in LT thin film matches, within experimental error, with the reported FTIR value (3476cm$^{-1}$) in LT crystals.

According to theoretical calculations of electronic structure by Inbar and Cohen, oxygen 2p states are contributing to energy levels at the top of the valence band [19]. Excitation of an electron from O-2p level in the valence band to the conduction band will give rise to coupling of hole to phonons of O-H/O-D vibrations. The optical and acoustic phonons of lattice and O-H/O-D vibrations absorb difference between the incident energy and the scattered energy. Energy conservation requires that

$$E_i = E_s \pm l\hbar\omega_{lattice} + m\hbar\omega_{OH} + n\hbar\omega_{OD} \qquad \ldots\ldots\ldots\ldots \quad (1)$$

Here, $E_i$ and $E_s$ are energies of incident and scattered photons respectively, l, m and n are whole numbers, $\hbar\omega_{lattice}$, $\hbar\omega_{OH}$ and $\hbar\omega_{OD}$ are energies of lattice phonons, vibration/ libration modes of O-H and vibration / libration modes of O-D respectively. The dispersion of RRS peaks of the order of 800cm$^{-1}$/eV can be explained if some energy is absorbed by lattice vibrations along with O-H and O-D vibrations. As shown in fig 5(a), the RRS peak positions of OH fundamental stretching mode are more (less) than the FTIR value for wavelengths lower (higher) than 250nm. The plus sign with the term $\hbar\omega_{lattice}$, in equation (1), is for incident wavelengths less than 250nm and negative for incident wavelengths more than 250nm.

Double resonance Raman scattering (DRRS) model has been used to explain the dispersion of the D band in graphite materials [16, 17]. The dispersion of the D band is minimum in crystalline graphite and maximum in amorphous graphite [16,17]. Whether,

DRRS model can be used to explain the dispersion of RRS of O-H stretching mode and higher harmonics in LT thin films, is a theoretical problem for further work. Initial measurements on LT crystals have shown dispersions of O-H stretching mode and its third harmonic similar to polycrystalline thin films. Therefore polycrystalline nature is not the cause of the dispersion.

**4. Conclusion**

In conclusion, the presence of OH /OD defects in LiTaO$_3$ thin films (without extra proton and deuteron exchange) has been confirmed from the resonant Raman spectrum. Resonant Raman peaks related with stretching and libration modes of O-H and O-D have been observed along with the band gap PL peak of LiTaO$_3$ thin films using UV xenon lamp source. The large dispersion of RRS peaks, with excitation energy is of the order of 800cm$^{-1}$/eV for fundamental and 2000 cm$^{-1}$/eV for third harmonic. Such large dispersion has not been reported in any other material. These observations of RRS peaks along with PL are important from basic and devices point of views.

**Acknowledgements**

The authors thank to Mr. C. Mukherjee for AFM and Dr. V. R. Reddy for grazing angle XRD characterization.

**Caption of figures:**

Fig.1: Photoluminescence excitation (PLE) spectra with the detector fixed at (1) 4.71eV and at (2) 4.42eV.

Fig.2: (a) Photoluminescence spectra (unpolarized) of polycrystalline $LiTaO_3$ thin film at room temperature. For curves 1 and 2 incident wavelengths are 240nm and 235nm respectively, (b) PL spectra with X-axis as shift energy with respect to incident energy.

Fig.3: Raman spectra of the polycrystalline $LiTaO_3$ thin film.

Fig.4: Bandgap PL spectra with superimposed Resonance Raman peaks of polycrystalline $LiTaO_3$ thin film at room temperature for incident wavelength 240nm at different angles of incidence.

Fig.5: (a) Peaks of O-H fundamental stretching mode, for constant incident photon flux, as a function of shift energy for different excitation wavelengths. Excitation wavelengths are written near respective peaks. (b) Intensity of maximum of O-H fundamental stretching mode peak, for constant incident photon flux, as a function of incident energy. (c) Curves 1 and 2 show the dispersion of the RR peaks related with O-H stretching mode and its third harmonic respectively as a function of the incident energy.

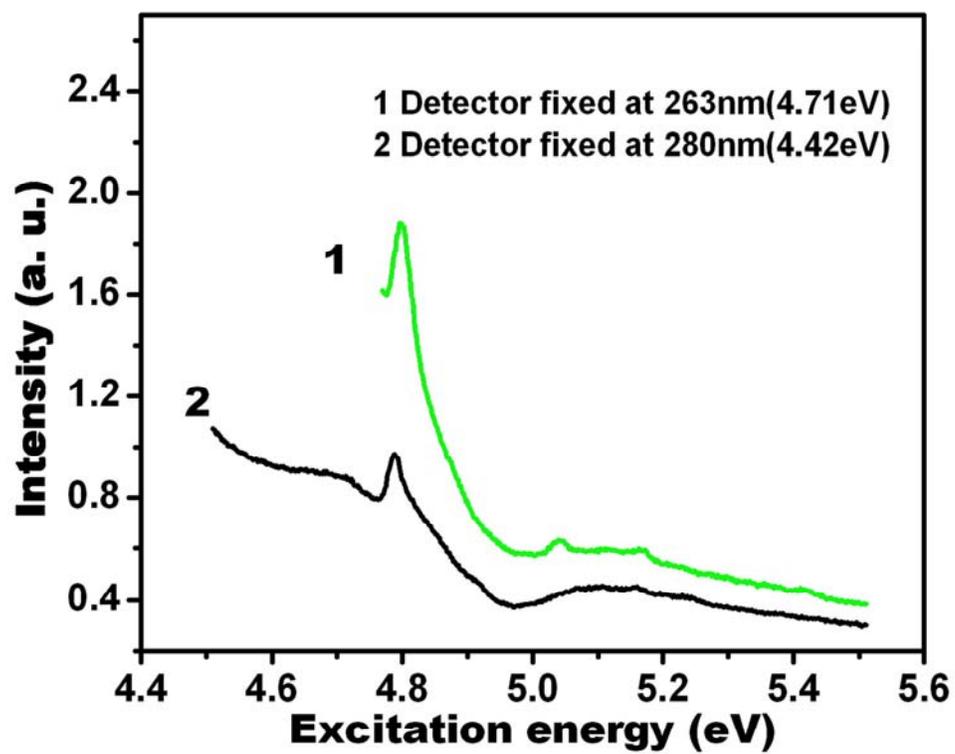

**Figure 1**

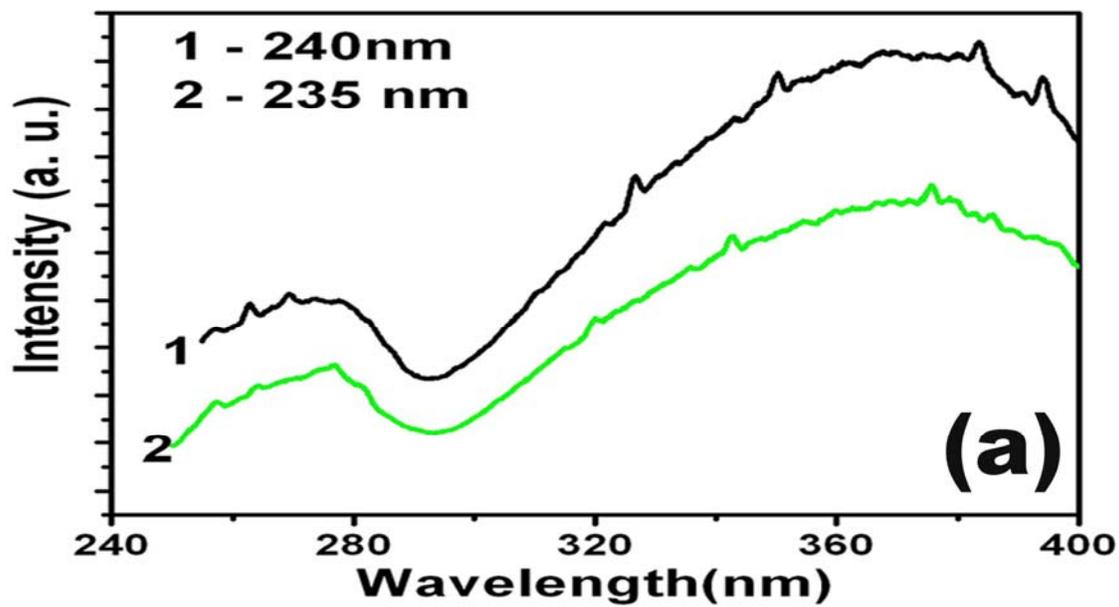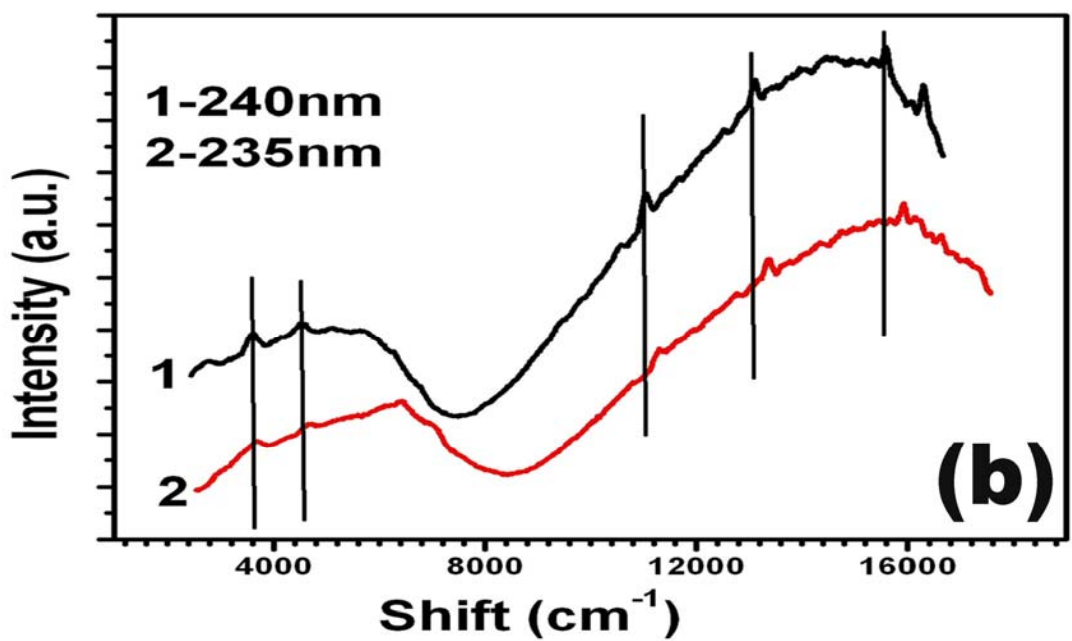

**Figure 2**

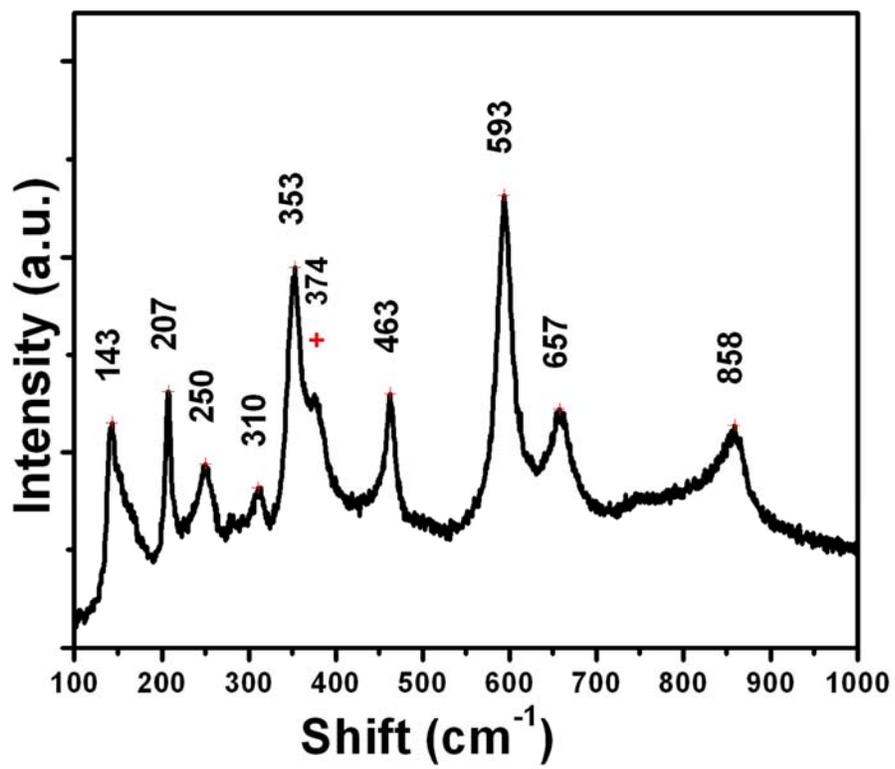

**Figure 3**

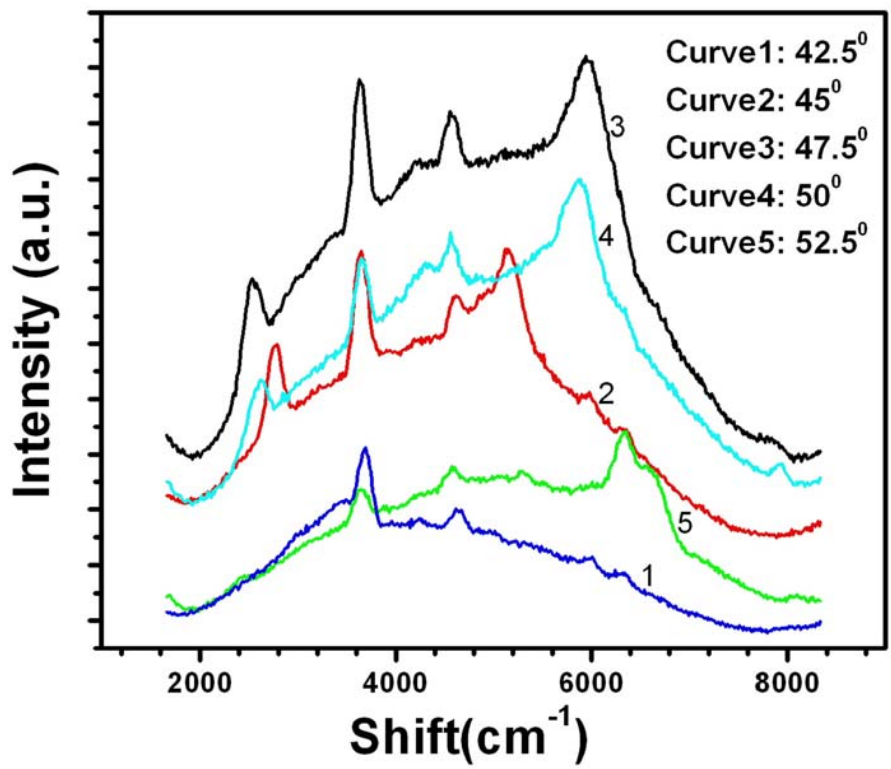

Figure 4

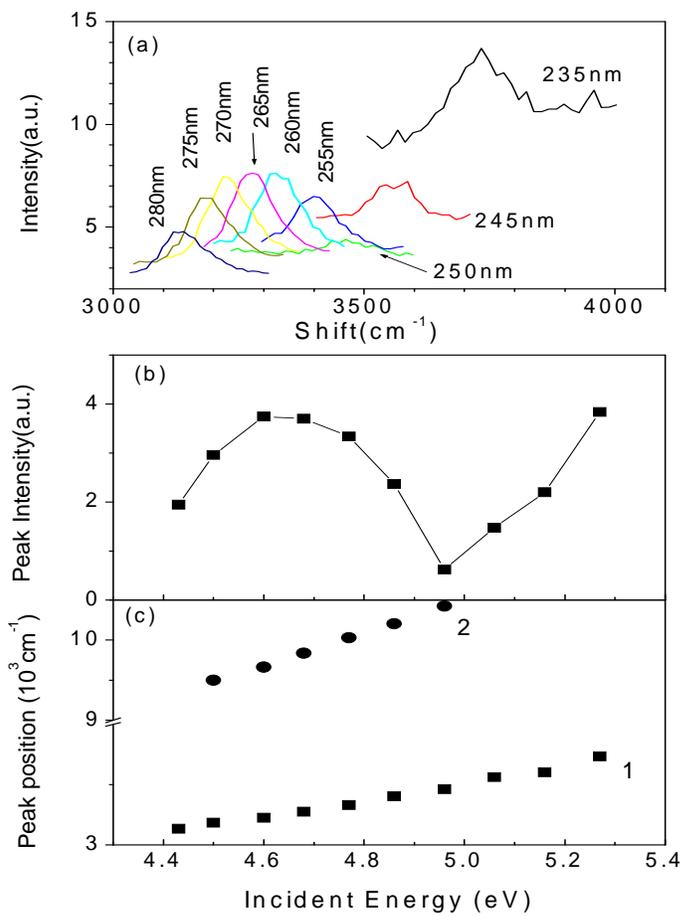

Figure:5

**Table 1**

Observed and reported wave numbers of the peak position due to OH /OD vibration, libration and combination of modes.

| Observed Peak positions (cm$^{-1}$) for excitation wavelength of 240nm. | FWHM (cm$^{-1}$) | Possible modes | Reported values (FTIR) of protonated and deutronated crystals (cm$^{-1}$) [4] |
|---|---|---|---|
| 2535 | 180 | Stretching of O-D | 2570 |
| 2764 | 180 | Stretching of O-H – libration of O-H | |
| 3637 | 180 | Stretching of O-H | 3476 |
| 4558 | 180 | Stretching + libration of O-H | 4449 |
| 5150 | 180 | 2$^{nd}$ harmonic of stretching of O-D | |
| 5975 | 346 | Combination of stretching modes of O-H with O-D; 2$^{nd}$ harmonic of stretching of O-D with libration of O-D | |
| 6350 | 200 | 2$^{nd}$ harmonic of stretching of O-H – libration of O-H | |
| 6630 | | 2 x (stretching + libration of O-D) | |
| 11040 | 200 | 3$^{rd}$ harmonic of stretching of O-H | |
| 13102 | 200 | 3$^{rd}$ harmonic of stretching of O-H with stretching of O-D | |
| 15595 | | 3$^{rd}$ harmonic of stretching of O-H with 2$^{nd}$ harmonic of stretching of O-D | |